\documentclass[12pt]{article}
\usepackage{graphicx}
\usepackage{amsfonts,amssymb,amsthm,amstext,amscd}

\newcommand{\beqa}{\begin{eqnarray}}
\newcommand{\eeqa}{\end{eqnarray}}

\newcommand{\be}{\begin{equation}}
\newcommand{\ee}{\end{equation}}
\newcommand{\beq}{\begin{equation}}
\newcommand{\eeq}{\end{equation}}
\newcommand{\bea}{\begin{eqnarray}}
\newcommand{\eea}{\end{eqnarray}}

\newcommand{\bear}{\begin{eqnarray}}
\newcommand{\eear}{\end{eqnarray}}
\begin{document}
\baselineskip=15.5pt
\pagestyle{plain}
\setcounter{page}{1}
\newfont{\namefont}{cmr10}
\newfont{\addfont}{cmti7 scaled 1440}
\newfont{\boldmathfont}{cmbx10}
\newfont{\headfontb}{cmbx10 scaled 1728}
\renewcommand{\theequation}{{\rm\thesection.\arabic{equation}}}
\font\cmss=cmss10 \font\cmsss=cmss10 at 7pt
\par\hfill{IFUP-TH/2013-25}
\vspace{1cm}

\begin{center}
{\huge{\bf  A novel cross-check of localization and non conformal holography}}
\end{center}

\vskip 10pt

\begin{center}
{\large Francesco Bigazzi$^{a}$, Aldo L. Cotrone$^{b}$, Luca Griguolo$^{c}$, Domenico Seminara$^{b}$}
\end{center}

\vskip 10pt
\begin{center}
\textit{$^a$ INFN, Sezione di Pisa; Largo B. Pontecorvo 3, I-56127 Pisa, Italy.}\\
\textit{$^b$ Dipartimento di Fisica e Astronomia, Universit\`a di
Firenze and INFN, Sezione di Firenze; Via G. Sansone 1, I-50019 Sesto Fiorentino
(Firenze), Italy.}\\
\textit{$^c$ Dipartimento di Fisica e Scienze della Terra, Universit\`a di Parma and INFN, Gruppo Collegato di Parma; Viale G.P. Usberti 7/A, 43100 Parma, Italy.}\\\vskip 10pt
{\small fbigazzi@pi.infn.it, cotrone@fi.infn.it, griguolo@fis.unipr.it, seminara@fi.infn.it}
\end{center}

\vspace{25pt}

\begin{center}
 \textbf{Abstract}
\end{center}

\noindent We precisely reproduce the perimeter law obeyed by Wilson loops on large spatial contours in planar ${\cal N}=2$ SYM at strong coupling, as recently deduced using localization, by means of a dual holographic model. The relevant supergravity background is sourced by $D5$-branes wrapped on a two-sphere in a Calabi-Yau two-manifold. Thus, localization and holography are cross-checked, for the first time, in a non conformal context where the gravity background is not asymptotically Anti de Sitter and the dual gauge theory has a logarithmically running coupling. We also notice that the same cross-check can be performed considering an alternative holographic description of ${\cal N}=2$ SYM based on a background sourced by fractional $D3$-branes.
\newpage

\section{Introduction}
\setcounter{equation}{0}
It has been recently argued in \cite{rz} (see also \cite{rz2,rz3}) that supersymmetric Wilson loops in some planar ${\cal N}=2$ supersymmetric gauge theories at strong coupling obey the perimeter law
\be
\log W[{\cal C}] = \mu L\,, \qquad (\mu L\gg 1)\,,
\label{pl}
\ee
where $L$ is the length of a large spatial contour and $\mu$, whose precise meaning will be clear in a moment, is proportional to the dynamical scale $\Lambda$ of the theory. The Wilson loops are taken in the fundamental representation and the above result has been derived both for ${\cal N}=2^*$ and for pure ${\cal N}=2$ Super-Yang-Mills (SYM) theories, using supersymmetric localization techniques \cite{pestun}.

The strategy goes as follows. One first considers the gauge theories compactified on a four-sphere $S^4$ in such a way that supersymmetry remains unbroken. In this case the partition function reduces to a finite dimensional integral over a real section of the complex moduli space of the theory in the Coulomb branch.\footnote{Notice that in \cite{rz}, by means of an explicit one-instanton computation, it was argued that instanton contributions to the partition function are suppressed in the large $N$ limit for both ${\cal N}=2$ and ${\cal N}=2^*$ SYM.} This in turn allows to compute relevant observables like the VEV of the Wilson loop on the big circle of $S^4$ (of length $L$) 
\be
W[{\cal C}_{\rm circle}] = \langle \frac{1}{N} {\rm Tr}\Bigl[P {\rm exp}\int_{{\cal C}_{\rm circle}} ds (i {\dot x}^{\mu}A_{\mu} + |\dot x| \Phi_0)\Bigr]\rangle\,,
\label{wilc}
\ee
where $\Phi_0$ is one of the two real adjoint scalar fields which combine to give rise to the complex field $\Phi$ of the ${\cal N}=2$ vector multiplet (e.g. $\Phi_0= {\rm Re}[\Phi]$). Localization amounts to replacing the fields $A_{\mu}$, $\Phi_0$ by their classical values $A_{\mu}=0$, $\Phi_0={\rm diag}(a_1,\dots, a_N)$ and performing the matrix integral over the classical configurations with an appropriate measure (that encodes the relevant dynamics of the particular theory). In the large $N$ limit, where the integral which defines the partition function is governed by a saddle point, the VEVs of the field $\Phi_0$ form a continuous distribution described by a density $\rho(a)$ which solves the saddle point equations. The eigenvualues are distributed on a interval $[-\mu, \mu]$ of the real axis, where their density is normalized to one
\be
\int_{-\mu}^{\mu}\rho(a) da =1\,.
\label{rel}
\ee  
Once $\rho(a)$ is determined, the circular Wilson loop is computed as
\be
W[{\cal C}_{\rm circle}] = \int_{-\mu}^{\mu} \rho(a) e^{L\, a}da\,.
\ee
The result for the Wilson loop in flat space is then obtained in the decompactification limit $\mu L\rightarrow\infty$. 

For ${\cal N}=2^*$ SYM, which is asymptotically conformal in the UV, the density distribution selected by localization precisely coincides, in the strong 't Hooft coupling regime, with the semi-circular one corresponding to the dual supergravity solution found in \cite{pw}. A holographic computation of large Wilson loops in that model, performed in \cite{brz}, shows that the perimeter law (\ref{pl}) is precisely reproduced. 

In the large $N$ limit of pure ${\cal N}=2$ SYM, when $\Lambda L$ is taken to be very large, it was shown in \cite{rz}  (see also \cite{rz2}) that the density distribution selected by localization is of the form
\be
\rho(a) = \frac{1}{\pi\sqrt{\mu^2-a^2}}\,,
\label{den}
\ee
where, adopting the renormalization scheme used in \cite{rz2}\footnote{According to \cite{rz2}, given a UV cutoff $M$, $\Lambda = M e^{- 4\pi^2/\lambda}$, where $\lambda$ is the 't Hooft coupling at the cutoff. This is the scheme we will use in the following. Notice that $\Lambda$ in \cite{rz2} differs by that in \cite{rz} by a factor $e^{-1-\gamma}$, where $\gamma$ is the Euler number.}  
\be
\mu = 2\Lambda\,.
\ee
A distribution of VEVs of this kind, for the theory in flat space-time, is found to describe the ${\mathbb Z}_2\subset U(1)_R$ symmetric points in the Coulomb branch of the theory where all types of monopoles become massless \cite{ds} (interestingly, these are the relevant vacua to consider in order to flow from the mass-deformed ${\cal N}=2$ theory to ${\cal N}=1$ SYM).

Crucially, the latter has a known dual supergravity description which has been found in \cite{bcz} (see also \cite{rev}).  The relevant type IIB supergravity background is sourced by $N$ $D5$-branes wrapping a two-cycle in a two-complex dimensional Calabi-Yau space ($CY_2$). The low energy dynamics of such $D5$-branes is described by a four-dimensional ${\cal N}=2$ SYM theory in various points of the Coulomb branch. The class of relevant solutions corresponding to the ${\mathbb Z}_2$-symmetric vacuum described above is parameterized by an integration constant $b$ which is precisely related to the ratio $\mu/\Lambda$.

Focusing on this class of solutions we show, by means of a simple holographic computation, that the perimeter law (\ref{pl}) for large spatial Wilson loops is precisely reproduced in the pure ${\cal N}=2$ SYM case. To our knowledge, this is the first time that localization and holography are cross-checked in a non conformal context where the gravity background is not asymptotically Anti de Sitter and the dual gauge theory has a logarithmically running coupling.

Finally, we notice that the above cross-check is realized also by considering an alternative - though not fully explicit - realization of the above mentioned ${\mathbb Z}_2$-symmetric vacuum, found in \cite{stefano} considering fractional $D3$-brane solutions. We will discuss this issue in the final part of this note. 
\section{The wrapped $D5$-brane background} 
\setcounter{equation}{0}
In this section we review the relevant gravity background, discussing along the way the crucial identification of the location of the holographic Wilson loop in the internal space, which allows to derive exactly formula (\ref{pl}) with a simple holographic calculation. 
The low energy dynamics of $N$ $D5$-branes suitably wrapped on a two-cycle in a $CY_2$, as shown in \cite{gauntlett,bcz}, is described by a four-dimensional ${\cal N}=2$ SYM theory. The dual supergravity description of the theory in the planar, strong coupling regime is, according to the holographic correspondence, given by the background which is sourced by the $D5$-branes. 

The ten-dimensional background which is relevant for the purpose of this note has been found in \cite{bcz}. It includes a six-form RR potential $C_6$ (whose full expression is not required now) and a string-frame metric and dilaton given by 
\bea
ds^2 &=& e^{\Phi_D} \left[dx_{\mu}dx^{\mu} + R^2 u (d\theta^2+\sin^2\theta d\varphi^2) + R^2 e^{2\lambda_1-\lambda_2-\lambda_3}du^2\right]+ R^2 e^{-\Phi_D}ds_3^2\,,\nonumber \\
ds_3^2 &=& e^{-6\lambda_1-2\lambda_2-2\lambda_3}\left[d\mu_1^2+d\mu_2^2+\cos^2\theta(\mu_1^2+\mu_2^2)d\varphi^2 -2\cos\theta(\mu_1 d\mu_2 - \mu_2 d\mu_1)d\varphi\right] + \nonumber \\
&& + e^{-4\lambda_1-2\lambda_2-2\lambda_3}\left[e^{-2\lambda_2}d\mu_3^2+e^{-2\lambda_3}d\mu_4^2\right]\,,\nonumber \\
e^{2\Phi_D} &=& \Delta\, e^{-3\lambda_2-3\lambda_3-6\lambda_1}\,,
\eea
where $u$ is a dimensionless radial variable (holographically related to the RG scale of the gauge theory) and
\bea
&&\Delta = e^{2\lambda_1}(\mu_1^2 + \mu_2^2) + e^{2\lambda_2}\mu_3^2+e^{2\lambda_3}\mu_4^2\,,\qquad R\equiv \sqrt{g_s N\alpha'}\,,\nonumber \\
&&\mu_{1,2}=\cos\theta'(\cos\phi_1, \sin\phi_1)\,,\quad \mu_{3,4}=\sin\theta'(\cos\phi_2,\sin\phi_2)\,.
\eea
The angular coordinates $\theta\in[0,\pi]$ and $\varphi\in[0,2\pi]$ describe the two-sphere which is wrapped by the $N$ $D5$-branes. The remaining ones are related to the transverse three-sphere and have ranges given by $0\le\theta'\le\pi/2, 0\le\phi_1,\phi_2\le2\pi$.
 
Finally, the functions $\lambda_{i}$ run with the radial variable $u$, according to the following relations
\bea
&&e^{\frac{\lambda_2+\lambda_3}{2}-\lambda_1} = \sqrt{\frac{e^{4u}+b^4}{e^{4u}-b^4}-\frac{1}{2u}+\frac{(1-b^4)e^{2u}}{2u(e^{4u}-b^4)}}\,,\nonumber \\
&&e^{\frac{\lambda_2+\lambda_3}{2}+\lambda_1}= \left(\frac{e^{2u}}{e^{4u}-b^4}\right)^{1/5}\left[\frac{e^{4u}+b^4}{e^{4u}-b^4}-\frac{1}{2u}+\frac{(1-b^4)e^{2u}}{2u(e^{4u}-b^4)}\right]^{-1/10}\,,\nonumber \\
&&e^{\lambda_2-\lambda_3} = \frac{e^{2u}-b^2}{e^{2u}+b^2}\,.
\label{lambdas}
\eea
Here $b$ is an integration constant.\footnote{The expressions (\ref{lambdas}) actually describe a subclass of the whole family of solutions found in \cite{bcz,rev} which in general depend on another integration constant. We focus on that subclass since, as we will show in the following, the relevant solution to be used in order to compare with the localization results turns out to belong to it.} The minimal allowed value of $u$ (i.e. the value $u_{min}$ at which the dilaton $e^{\Phi_D}$ goes to zero and the Ricci scalar diverges) generically depends on $b$ and on the angular directions. 

When $u\rightarrow\infty$ (which corresponds to the UV of the dual gauge theory) the solution asymptotes to a linear dilaton background. When $u\rightarrow u_{min}$ (the IR) the solution has a ``good" singularity according to the criteria in \cite{goods}.

If $b=0$ one gets $\lambda_2=\lambda_3$. In this case the metric has two $U(1)$ isometries corresponding to shifts of the angles $\phi_1$ and $\phi_2$. These correspond respectively to the classical $U(1)_J\subset SU(2)_R$ and $U(1)_R$ global symmetries of the dual field theory. In the general case with $b\neq0$  the $U(1)_R$ symmetry is broken to ${\mathbb Z}_2$. As it has been discussed in \cite{bcz,Babington:2001nh,Apreda:2001qb,rev}, the scalar fields $\lambda_2+\lambda_3$ and $\lambda_2-\lambda_3$ are in fact dual to the operators $Tr\Phi{\bar\Phi}$ and $Tr\Phi^2$, respectively.

The classically $U(1)_R$ symmetric solution, which was also independently found in \cite{gauntlett} (see also \cite{dlm}), was argued to correspond to a point of the Coulomb branch of the dual field theory where the VEVs of the scalar field are spherically distributed. The $U(1)_R$ symmetry, which is actually broken to ${\mathbb Z}_{4N}$ by the anomaly,\footnote{This breaking shows up on the gravity side by examining the flux of the RR potential $C_2$ through $S^2$, see e.g. \cite{dlm}.} acts as a ${\mathbb Z}_{2N}$ subgroup on that point of the Coulomb branch. This is not the solution to be considered in order to match the localization results found in \cite{rz}. We will thus consider just the $b\neq0$ solution from here on.

A standard procedure allows to understand which point of the Coulomb branch of the ${\cal N}=2$ SYM theory corresponds to a given dual supergravity solution. First, one considers the symmetry breaking pattern $SU(N+1)\rightarrow SU(N)\times U(1)$, which is induced (at the classical level) when one eigenvalue $\phi$ of the adjoint scalar field $\Phi$ gets a non-zero VEV: it can be realized moving one of the $N+1$ wrapped $D5$-branes from the original stack into the transverse space. If the remaining stack of $N$ branes describes the theory on a generic point of the Coulomb branch, the corresponding $SU(N)$ symmetry is actually broken to $U(1)^{N-1}$. All in all, the adjoint scalar field of the $SU(N+1)$ ${\cal N}=2$ SYM theory will thus take VEVs according to
\be
\Phi= {\rm diag}\left(\phi, a_1 -\frac{\phi}{N}, a_2-\frac{\phi}{N}, \dots , a_N -\frac{\phi}{N}\right)\,.
\label{Phi}
\ee
The $D5$-brane probe action will capture the effective description for the corresponding $U(1)$ factor and the scalar field $\phi$. The moduli space will be then identified with the two-dimensional domain in the transverse space where the $D5$-brane (which will then probe the background sourced by the remaining stack) can move freely.  As it has been shown in \cite{bcz}, in the setup introduced above, this (no-force, BPS) condition is realized when $\theta'=\pi/2$. The corresponding subspace has metric given by
\be
ds^2|_{\theta'=\frac{\pi}{2}}= e^{\Phi_D} \left[dx_{\mu}dx^{\mu} + R^2 u (d\theta^2+\sin^2\theta d\varphi^2) \right] + e^{-\Phi_D} R^2 dw\,d{\bar w}\,,
\label{slice}
\ee
where we have introduced the complex coordinate
\be
w = e^{u+i\phi_2} + b^2 e^{-u-i\phi_2}\,,
\ee
and it is understood that the dilaton is evaluated at $\theta'=\pi/2$. It is the plane spanned by the complex coordinate $w$ - which is transverse to the $D5$-brane probe (whose worldvolume is along $x^{\mu}, \theta, \varphi$) - which is mapped to the moduli space. 

The $D5$-brane action in static gauge is given by
\be
S_{D5} = -T_5\int d^4x d\theta d\varphi e^{-\Phi_D}\sqrt{-\det [G + 2\pi\alpha' F]} + T_5\int\left( C_6 + \frac{1}{2} C_2\wedge (2\pi\alpha') F\wedge (2\pi\alpha')F\right)\,,
\ee  
where $G_{ab}$ is the induced metric, $F$ is the $U(1)$ gauge field strength on the brane,
\be
T_5 = \frac{1}{(2\pi)^5\alpha'^3 g_s}
\ee
is the $D5$-brane tension and the relevant part of $C_6$ is given by
\be
C_6 = R^2 e^{2\Phi_D} u\, dx_0\wedge dx_1\wedge dx_2\wedge dx_3\wedge d\theta \wedge  \sin\theta d\varphi\,.
\ee
The two-form potential $C_2$ is obtained from $C_6$ by Hodge duality.

Computing the above action on the $\theta'=\pi/2$ subspace of the background it is easy to see that no potential term is induced. After integrating over the two sphere, in the low-energy limit (i.e. up to quadratic terms in derivatives), the action reduces to\footnote{It is worth noting that the dilaton factors drop out.}
\be 
S = \frac{1}{4\pi}\int d^4x\left[-\frac{1}{2}{\rm Im}(\tau(\phi)) F^2 -{\rm Im}(\tau(\phi))\partial\phi \partial {\bar \phi} +\frac{1}{2}{\rm Re}(\tau(\phi)) F{\tilde F}\right]\,.
\ee
This is precisely the effective action for the $U(1)$ gauge field (with field strength $F$) and the complex scalar $\phi$ we were looking for. The complex coupling $\tau(\phi)$ is defined as usual $\tau = (\theta/2\pi) + i (4\pi/g^2)$ and the map between gauge and gravity objects is given by
\be
\tau = i\, \frac{N}{\pi}\left[\cosh^{-1}\left(\frac{w}{2b}\right) + \log b\right]\,,\qquad w = \frac{2\pi\alpha'}{R}\phi = \frac{2\pi\sqrt{\alpha'}}{\sqrt{g_s N}}\,\phi\,,
\label{maps}
\ee 
or, in terms of the coordinates $u,\phi_2$, 
\be
{\rm Im}(\tau) = \frac{N}{\pi}u\,,\quad  {\rm Re}(\tau) = - \frac{N}{\pi}\phi_2\,.
\ee
Notice that the complex coupling $\tau$ has a branch cut along the interval $[-2b, 2b]$ on the real axis.

Now, on the field theory side we know that the complex coupling can be obtained from the prepotential
\be
{\cal F}= \frac{i}{4\pi}\sum_{i<j} (a_i-a_j)^2\log\frac{(a_i-a_j)^2}{e^3\Lambda^2}\,,
\label{prepo}
\ee
which is one-loop exact.\footnote{Instanton effects are neglected according to the fact that we only work at the level of classical supergravity backgrounds on the dual holographic side. It would be interesting to better investigate the issue of instantonic corrections for the class of vacua we are focusing on. As it was mentioned in the introduction, those corrections have been argued to be subleading at large $N$ on the vacuum selected by localization \cite{rz}.} Here $\Lambda$ is the dynamical scale of the theory in the renormalization scheme of e.g. \cite{ds}. A simple computation given in the Appendix, confirms that this is precisely the scheme used in \cite{rz2}. 

Considering a distribution of eigenvalues as in (\ref{Phi}) we can thus deduce that
\be
\tau(\phi) = \frac{\partial^2{\cal F}}{\partial\phi^2}=\frac{i}{\pi}\sum_{i} \log\frac{(\phi-a_i)}{\Lambda}\,.
\label{taup}
\ee
In the large $N$ limit, the sum above can be replaced by an integral, so that we can write
\be
\tau(\phi) =\frac{i N}{\pi} \int d^2a\, \rho(a) \log\frac{(\phi-a)}{\Lambda}\,,
\ee
by means of a unit-normalized density distribution $\rho(a)$ of the complex VEVs. 

From the explicit expression found in (\ref{maps}) we can thus deduce \cite{bcz} that the supergravity solution here considered corresponds to a point of the Coulomb branch where the density distribution of the VEVs of the adjoint scalar field is precisely of the same form given in (\ref{den}), with 
\be
\mu = 2b\,\Lambda\,,\qquad \Lambda=\frac{\sqrt{g_s N}}{2\pi \sqrt{\alpha'}}\,,
\label{rela}
\ee
which follow from the map between $\phi$ and $w$ given in (\ref{maps}).  As we have noticed above, the distribution of VEVs which is selected by localization has $\mu=2\Lambda$ in the scheme used in \cite{rz2}. In order to see whether the holographic computations (e.g. of the Wilson loop) match with the field theory results obtained using localization techniques, we thus need to focus on the particular background with
\be
b=1\,.
\ee 
Notice, moreover, the crucial fact that the VEV distribution selected by localization has support on the real $\phi_2=0$ (mod $\pi$) slice of the moduli space in our gravity setup. In fact, precisely this distribution is the one which solves the matrix model which emerges from localization. This, together with (\ref{maps}) and (\ref{rela}), is one of the essential ingredients to consider when holographically computing the Wilson loops which are relevant in the framework of localization.
\section{The holographic Wilson loop}
\setcounter{equation}{0}
We want to focus on Wilson loops of the form given in (\ref{wilc}). We will consider general spatial contours in the strict $L\mu\rightarrow\infty$ limit. In this case any contour can be approximated by an interval of infinite extension. Holographically, the Wilson loop is computed by means of the formula
\be
\log W[{\cal C}] = - S_{NG}^{r}\,,
\label{holw}
\ee
where $S_{NG}^r$ is the renormalized Euclidean Nambu-Goto action for an open string which is attached to the contour ${\cal C}$ on the boundary.

The point in the internal manifold on which the open string sits is determined by the scalar couplings of the Wilson loop.  The supersymmetric Wilson loop (\ref{wilc}) couples to the adjoint scalar field of the ${\cal N}=2$ theory which gets a VEV on a real slice of the Coulomb branch (according to the fact that the relevant large $N$ solution of the matrix model obtained through localization has support on the real axis). In our setup, this implies considering the open string to be attached to a BPS probe brane at a precise fixed value of the angle $\phi_2$
\be
\theta'=\frac{\pi}{2}\,,\qquad \phi_2=0\,.
\ee
Once the position of the string in the internal space is determined, the actual holographic calculation is straightforward.
The open string embedding relevant for the Wilson loop computation in the setup described in the previous section is  described as follows
\be
\tau = x \in[-\frac{L}{2},\frac{L}{2}]\,,\qquad \sigma = w|_{\phi_2=0}\in [2, M]\,,
\ee
where $M$ is an UV cutoff and in terms of the original radial variable $u$ (which has minimal value $u_{min}=0$ for the selected choices of angles and parameter $b$)
\be
w|_{\phi_2=0} = e^u + e^{-u}\,,
\ee
where we have fixed $b=1$ for the reasons explained above. From eq. (\ref{slice}) it is immediate to see that the induced metric on the open string world-sheet reads
\be
ds_2^2 = e^{\Phi_D} dx^2 + e^{-\Phi_D} R^2 (dw|_{\phi_2=0})^2\,.
\ee
The Euclidean Nambu-Goto action reads
\be
S_{NG} = \frac{1}{2\pi\alpha'} \int d\tau d\sigma \sqrt{\det g_{2}} = \frac{L}{2\pi\alpha'}R\int_{2}^{M} dw|_{\phi_2=0} = L \frac{\sqrt{g_s N}}{2\pi\sqrt{\alpha'}} \left[M - 2\right] = L\left[\phi[M] - \mu\right]\,,
\ee
where we have used the gauge/gravity dictionary introduced above. After subtracting the UV divergent term $L\,\phi[M]$ and using the holographic relation (\ref{holw}) we get the perimeter law
\be
\log W[{\cal C}] = \mu L\,,
\ee
precisely reproducing the field theory result found in \cite{rz}. 

It would be interesting to further cross-check holography and localization beyond the leading order decompactification limit. This would amount on finding the supergravity dual of ${\cal N}=2$ SYM on $S^4$ by studying a suitable deformation of the wrapped $D5$-background described above, along the lines of what has been recently done in the ${\cal N}=2^*$ case in \cite{free}. This would allow us to also holographically compute other relevant observables, like the field theory free energy on $S^4$, which have been also determined on field theory grounds using localization in \cite{rz}.
\section{The fractional $D3$-brane setup}
\setcounter{equation}{0}
The same ${\mathbb Z}_2$-symmetric distribution of VEVs on which we have focused above, was realized in \cite{stefano} in a context in which ${\cal N}=2$ SYM is obtained as the low energy limit of $N$ fractional $D3$-branes at the ${\mathbb C}\times {\mathbb C}^2/{\mathbb Z}_2$ orbifold singularity.
In this case the 10d string-frame metric has a form given by (see e.g. \cite{grana,matteo}) 
\be
ds_{10}^2 = e^{\frac{\Phi_D}{2}}\left[H^{-1/2} dx_{\mu}dx^{\mu}+ H^{1/2} e^{-\Phi_D} dz d{\bar z} + H^{1/2} (dw_1 d{\bar w}_1+ dw_2 d{\bar w}_2)\right]\,.
\ee
The orbifold action acts on the transverse complex coordinates as $w_1\rightarrow -w_1$, $w_2\rightarrow - w_2$ leaving $z$ unchanged. The dilaton $\Phi_D$ is actually constant if the background is sourced by just $N$ fractional $D3$-branes.\footnote{Other setups, relevant for describing SYM theories coupled with matter hypermultiplets, also include $D7$-branes extended along $x^{\mu}, w_1, w_2$. In those cases the dilaton is a non trivial function of the complex coordinate $z$.} The warp factor $H$ depends on the transverse coordinates in a non-trivial way which depends on the field theory vacuum to which the background is dual. In the case we are interested in, the function $H$ has not been explicitly computed. However, as we will see in a moment, knowledge of $H$ is not necessary for the holographic computation of the Wilson loop we are focusing on. 

A probe analysis analogous to the one considered for the wrapped $D5$-brane case allows to identify the plane spanned by $z$ with the moduli space of the theory. The relation between the complex scalar field VEV $\phi$ and $z$ is given by
\be
\phi= \frac{z}{2\pi\alpha'}\,.
\label{relaz}
\ee
The crucial features of the supergravity solution sourced by the fractional branes are captured by the value of the twisted sector scalar field
\be
\gamma = c + \tau b = \frac{1}{4\pi^2\alpha' g_s}\int \left( C_2 + \tau B_2\right)\,,\quad \tau = C_0 + i e^{-\Phi_D}\,,
\ee
where $C_0, C_2$ are RR potentials, $B_2$ is the NSNS antisymmetric field and the integral is done over the vanishing two-cycle of the orbifold. The explicit expression of the warp factor can be determined by solving a differential equation depending on $\gamma$.

In \cite{stefano} it was shown that there is a class of fractional $D3$-brane solutions which is a good candidate (alternative to the wrapped $D5$-brane one) to provide a dual description to the ${\cal N}=2$ SYM ${\mathbb Z}_2$-symmetric vacuum we are interested in. The twisted sector scalar field (which actually gives the complex coupling of the $U(1)$ theory on the probe brane\footnote{See \cite{lerdato} for a detailed discussion on the matter.}) reads in fact
\be
\gamma = \frac{i N}{\pi} \cosh^{-1} \left(\frac{v}{2}\right)= \frac{i N}{\pi} \cosh^{-1}\left( \frac{\phi}{\mu}\right)\,,
\ee
where we have used the relation $z = 2\pi\alpha' \Lambda v$ given in \cite{stefano} together with formula (\ref{relaz}) and we have defined
\be
\mu = 2\Lambda\,,
\ee
where $\Lambda$ is the dynamical scale of the theory. Notice the matching between these expressions and the corresponding ones in the wrapped D5-brane context with $b=1$.

We can now repeat almost literally the holographic Wilson loop computation considered in the previous section. 
The relevant open string embedding is
\be
\tau = x \in [-L/2,L/2]\,,\quad \sigma = \zeta \equiv {\rm Re}[z]\in[ 2\pi\alpha'\mu, \infty]\,,
\ee
and the string is attached to a BPS probe at $w_1, w_2=0$ and ${\rm Im[z]}=0$. The world-sheet metric then reads
\be
ds_2^2 = e^{\frac{\Phi_D}{2}}H^{-1/2} dx^2 + e^{-\frac{\Phi_D}{2}}H^{1/2} d\zeta^2\,,
\ee
so that the on-shell Nambu-Goto action reads
\be
S_{NG} = \frac{L}{2\pi\alpha'} \int_{ 2\pi\alpha'\mu}^{2\pi\alpha' M} d\zeta = L M - L\mu\,,
\ee
where $M$ is as usual an UV cutoff. After subtracting the perimeter divergence $LM$ one gets a renormalized Nambu-Goto action which in turn gives rise to the expected relation
\be
\log W = L \mu\,,
\ee 
for the Wilson loop VEV on a large spatial contour.

It could be interesting to consider whether analogous matchings with localization computations are realized for ${\cal N}=2$ SYM theories coupled with fundamental hypermultiplets. In this case one should explore solutions involving fractional $D7$-branes too, as in \cite{grana,matteo}. 
\vskip 15pt \centerline{\bf Acknowledgments} \vskip 10pt \noindent We are grateful to Konstantin Zarembo for illuminating discussions and relevant comments and to Stefano Cremonesi for signaling us his relevant works on the fractional $D3$-brane setup and for useful comments. We thank the Galileo Galilei Institute for Theoretical Physics, Florence for hospitality during the completion of this work. 
\appendix
\setcounter{equation}{0}
\section{On the renormalization scheme in \cite{rz2}}\label{app.rensche}
A simple way to realize that the renormalization scheme in \cite{rz2} actually coincides with the one adopted in (\ref{prepo}), (\ref{taup}) is as follows.\footnote{We thank K. Zarembo for suggesting this check to us.} The partition function for large $N$, ${\cal N}=2$ SYM on $S^4$ as can be deduced from eq. (5.4) in \cite{rz2} (with $N_f=0$) reads
\be
Z = \int d^{N-1}a\, \Pi_{i<j}(a_i-a_j)^2 H^{2}(a_i-a_j) e^{2N (\log\Lambda + \gamma +1)\sum_i a_i^2}\,,
\label{Zeta}
\ee
where $\gamma$ is the Euler number, the instanton contribution ${\cal Z}_{inst}$ has been put to one according to the observations in \cite{rz} and the radius $r$ of $S^4$ has been set to one. To perform the decompactification limit we restore the $r-$dependence and we take $r a_i\gg1$, then the function $H(x)$ behaves to leading order as (see e.g. eq. (A.4) in \cite{rz2})
\be 
\log H(x) = -\frac12 x^2 \log x^2 +\left(\frac12-\gamma\right)x^2 + {\cal O}(\log x^2)\,.
\ee
Taking, in this limit, a VEV distribution as in (\ref{Phi}) and considering for instance the case $\phi\gg a_j$ just to pick up the UV behavior, it turns out that the $\phi$-dependent factor in the integrand of (\ref{Zeta}) goes, to leading order, as
\be
{\cal Z}[\phi] \rightarrow e^{ 4\pi i  r^2{\cal F}[\phi] }\,,
\ee
where
\be
{\cal F}[\phi] = \frac{i N}{4\pi} \phi^2 \log\frac{\phi^2}{e^3\Lambda^2}
\ee
precisely coincides  with the prepotential (\ref{prepo}) in the same limit. 

\end{document}